\documentclass[pra,aps,twocolumn,twoside,english]{revtex4}

\usepackage{amssymb}
\usepackage{amsmath}
\usepackage{amscd}
\usepackage{babel}
\usepackage{latexsym}  
\usepackage{epsfig}
\usepackage{bbm}
\newtheorem{defi}{Definition}
\newtheorem{lemma}[defi]{Lemma}
\newtheorem{thm}[defi]{Theorem}

\newtheorem{rem}[defi]{Remark}
\newtheorem{prop}[defi]{Proposition}
\newtheorem{exempel}[defi]{Example}

\def\p{\pi}

\newcommand{\qed}{\hfill $\Box$}
\newcommand{\tr}{{\operatorname{Tr}}}
\newcommand{\id}{{\operatorname{id}}}
\newcommand{\supp}{{\operatorname{supp}\,}}
\newcommand{\bra}[1]{{\langle{#1}|}}
\newcommand{\ket}[1]{{|{#1}\rangle}}
\newcommand{\ketbra}[1]{{\ket{#1}\!\bra{#1}}}

\newcommand{\R}{{\mathbbm{R}}}

\newcommand{\E}{{\mathbbm{E}}}

\newcommand{\fset}[1]{{\mathcal{#1}}}

\newcommand{\1}{{\openone}}

\newenvironment{expl}[1][{}]{\begin{exempel} {#1}\normalfont}{\end{exempel}}

\newlength{\blank}
\newlength{\equalsign}
\newenvironment{beweis}[1][{\hspace{-\blank}}]{{\noindent\emph{Proof~{#1}.\ }}}{\hfill $\Box$\vskip 0.5\baselineskip}

\begin{document}

\title{Distillation of secret key and\protect\\  entanglement from quantum states}
\date{11th June 2003}
\author{Igor Devetak}
\email{devetak@us.ibm.com}
\affiliation{IBM T.~J.~Watson Research Center, PO Box 218, Yorktown Heights, NY 10598, USA}
\author{Andreas Winter}
\email{winter@cs.bris.ac.uk}
\affiliation{Department of Computer Science, University of Bristol, Merchant Venturers Building,\\ Woodland Road, Bristol BS8 1UB, United Kingdom}
\begin{abstract}
  We study and solve the problem of distilling secret key from quantum states
  representing correlation between two parties (Alice and Bob) and an eavesdropper
  (Eve) via one--way public discussion: we prove a coding theorem to achieve the
  ``wire--tapper'' bound, the difference of the mutual information Alice--Bob
  and that of Alice--Eve, for so--called cqq--correlations,
  via one--way public communication.
  This result yields information--theoretic formulas for the distillable secret key,
  giving ``ultimate'' key rate bounds if Eve is assumed to possess a purification
  of Alice and Bob's joint state.
  \par
  Specialising our protocol somewhat and making it coherent leads us to
  a protocol of entanglement distillation via one--way LOCC (local operations and classical
  communication) which is asymptotically optimal:
  in fact we prove the so--called ``hashing inequality''
  which says that the coherent information (i.e., the negative conditional
  von Neumann entropy) is an achievable EPR rate.
  This result is well--known to imply a whole set of distillation and capacity
  formulas which we briefly review.
\end{abstract}

\maketitle

\section{Introduction}
\label{sec:intro}
Entanglement and secret correlation share an ``exclusiveness'' --- in the one case
towards the total outside world, in the other towards an entity ``Eve'' --- that
has led quantum information scientists to speculate on a systematic relation between
their theories: the works in this direction range from building
analogies~\cite{collins:popescu} to using entanglement to prove information theoretic
security of quantum key distribution~\cite{shor:preskill},
to attempts to prove the equivalence of the distillability of secret key and of
entanglement~\cite{acin:key,bruss:et:al}.
\par
Of course there are also conceptual differences: while the task of
distilling secret perfect correlation derives from potential cryptographic
applications (and requires a third, malicious, party to formulate
the operational problem), entanglement is useful for simple transmission tasks
between two perfectly cooperating parties, as exemplified by dense
coding~\cite{Dense:coding} and teleportation~\cite{Teleportation}.
\par
The present paper falls into the third of the above categories,
for we address the two questions,
of distilling secret key from many copies of a quantum state (itself a generalisation
of classical information theoretic work begun by Maurer~\cite{maurer} and Ahlswede and
Csisz\'{a}r~\cite{AC:1}) by public discussion and of distilling EPR pairs by local
operations and classical communication (LOCC), in a unified way.
To be more precise,
after describing a protocol  for secret key distillation from a state
by one--way public discussion,
we show how secrecy codes of a particular structure can be converted into one--way
LOCC entanglement distillation protocols
achieving the \emph{coherent information}, as was conjectured for some time under the
name of the ``hashing inequality'' (after the hashing protocol in~\cite{BDSW}
which attains the bound for Bell--diagonal two--qubit states).
It is well--known from~\cite{HHH} that this inequality yields information theoretic
characterisations of distillable entanglement under general LOCC, as well as the quantum
transmission capacity, without, with forward and with bidirectional classical side channel
(the first of these capacity theorems proved recently by Shor~\cite{shor:Q}, 
following a heuristic argument of Lloyd~\cite{Lloyd:Q}, and subseqently 
in~\cite{devetak}).
Our approach is very close to that of~\cite{devetak}, and
--- as far as secret key distillation is concerned --- the work~\cite{cai:yeung}:
while here our resource is a three--party quantum state (``static'' model),
these papers deal with the ``dynamic'' analogue, where the resource is a 
quantum/wiretap channel.
\par
As for the structure of the paper: the main result of the cryptographic part is
theorem~\ref{thm:cqq:1way:coding} in section~\ref{sec:oneway-secret:cqq};
the form of the optimal rates is then not hard
to obtain, as we shall show in the detailed discussion. It is
theorem~\ref{thm:cqq:1way:coding} which we return to in the entanglement distillation
part: a very general modification of the coding procedure will give us
theorem~\ref{thm:hashing}, the hashing inequality; and as before, the form of
the optimal rates is not hard to get from there. A reader only interested in
entanglement distillation can thus skip the second part of
section~\ref{sec:oneway-secret:cqq}:
there the general form of optimal one--way distillable secret key is derived.
In section~\ref{sec:oneway-ent} we turn to one--way entanglement distillation,
proving the hashing inequality and exhibiting the general form of optimal
one--way distillation; then in section~\ref{sec:cap-and-ent} the consequences
of the hashing inequality are detailed.
Appendices collect the necessary facts about typical subspaces
(\ref{app:types}), some miscellaneous lemmas (\ref{app:facts})
and miscellaneous proofs (\ref{app:proofs}).

\section{One--way secret key distillation}
\label{sec:oneway-secret:cqq}
We will first study and solve the case of cqq--correlations, i.e., the initial state
$\rho^{ABE}$ has the form
\begin{equation}
  \label{eq:cqq}
  \rho^{ABE} = \sum_{x\in{\cal X}} P(x)\ketbra{x}^A\otimes\rho_x^{BE}.
\end{equation}
Then $n$ copies of that state can be written
$$\left(\rho^{ABE}\right)^{\otimes n}
        = \sum_{x^n} P^n(x^n)\ketbra{x^n}^A\otimes\rho^{BE}_{x^n},$$
with $x^n=x_1\ldots x_n$ and
\begin{align*}
  \ket{x^n}       &= \ket{x_1}\otimes\cdots\otimes\ket{x_n}, \\
  \rho^{BE}_{x^n} &= \rho^{BE}_{x_1}\otimes\cdots\otimes\rho^{BE}_{x_n}.
\end{align*}
Let $X$ be a random variable with distribition $P$, and corresponding to the
$n$ copies of $\rho$ consider independent identically distributed (i.i.d.)
realisations $X_1,\ldots,X_n$ of $X$.
\par
A \emph{one--way key distillation protocol} consists of:
\par
\begin{itemize}
  \item A channel $T:x^n\longrightarrow(\ell,m)$, with
    range $\ell\in\{1,\ldots,L\}$ and $m\in\{1,\ldots,M\}$.
  \item A POVM $D^{(\ell)}=(D^{(\ell)}_m)_{m=1}^M$ on $B^n$
    for every $\ell$.
\end{itemize}
\par
The idea is that Alice generates $T(X^n)=(\Lambda,K)$; her version of the key
is $K=m$, while she sends $\Lambda=\ell$ to Bob. He obtains his
$K'$ by measuring his system $B$ using $D^{(\ell)}$:
$$\Pr\{ K'=m|\Lambda=\ell,X^n=x^n \} = \tr\bigl( D^{(\ell)}_m \rho^B_{x^n} \bigr).$$
For technical reasons we assume that the communication has a rate
$L\leq 2^{nF}$, for some constant $F$.
\par
We call this an \emph{$(n,\epsilon)$--protocol} if
\par
\begin{enumerate}
  \item $\Pr\{ K\neq K' \} \leq \epsilon$.
  \item $\left\| {\rm Dist}(K)-\frac{1}{M}1_{\{1,\ldots,M\}} \right\|_1 \leq \epsilon$.
  \item There is a state $\sigma_0$ such that for all $m$,
    $$\left\| \sum_{x^n,\ell}\!\!\!
               \Pr\{X^n\!=\!x^n\!,\Lambda\!=\!\ell|K\!=\!m\}\ketbra{\ell}\otimes\rho^E_{x^n}\!-\sigma_0
       \right\|_1                                                                      \leq \epsilon.$$
\end{enumerate}
\par
We call $R$ an \emph{achievable rate} if for all $n$ there exist
$(n,\epsilon)$--protocols with $\epsilon\rightarrow 0$ and
$\frac{1}{n}\log M\rightarrow R$ as $n\rightarrow\infty$.
(The convention in this paper is that $\log$ and $\exp$ are understood
to be to basis $2$.) Finally define
$$K_\rightarrow(\rho):=\sup\{R:R\text{ achievable}\},$$
the \emph{one--way (or forward) secret key capacity of $\rho$}.
\par
Before we can formulate our first main result, we have to introduce some
information notation: for a quantum state $\rho$ we denote
the von Neumann entropy $H(\rho)=-\tr\rho\log\rho$, and the Shannon
entropy of a probability distribution $P$, $H(P)=-\sum_x P(x)\log P(x)$.
If the state is the reduced state of a multi--party state, like
the $\rho^{ABE}$ above, we write $H(A)=H(\rho^A)$, etc. In the particular
case of eq.~(\ref{eq:cqq}), obviously $H(\rho^A)=H(P)$.
For a general bipartite state $\rho^{AB}$ define the
\emph{(quantum) mutual information}
$$I(A:B)=H(A)+H(B)-H(AB),$$
which for the cq--state of eq.~(\ref{eq:cqq}) is easily checked to be equal to
$$H(\rho^B) - \sum_x P(x)H(\rho^B_x),$$
a quantity known as the \emph{Holevo bound}~\cite{holevo:bound}
and which we denote $I(P;\rho^B)$, reflecting in the notation
the distribution $P$ and the cq--channel with channel states $\rho^B_x$.
We shall often use the abbreviation $I(X;B)$ for this latter, if
the states and distribution of the random variable $X$ are implicitly
clear: this latter notation has the advantage that for any $U$
jointly distributed with $X$, $I(U;B)$ makes sense
immediately, without our having to write down a composite state.
\par
Finally, for a tripartite state $\rho^{ABC}$, define
the \emph{(quantum) conditional mutual information}
$$I(A:C|B) := H(AB)+H(BC)-H(ABC)-H(B),$$
which is non--negative by strong subadditivity~\cite{lieb:ruskai}.
Usually the state these notations refer too will be clear from the context;
where not we add it in subscript. Observe that for a classically correlated
system $B$, the conditional mutual information takes the form of
a probability average over mutual informations: e.g., for the
state of eq.~(\ref{eq:cqq}),
$$I(B:E|A) = \sum_x P(x)I(B:E)_{\rho_x}.$$
Also for conditional mutual information we make use of the hybrid
notation involving random variables: for example, for random variables
$T$ and $U$, jointly distributed with $X$, $I(U;B|T)$ is the average
over $T$ of Holevo quantities as above.
\par
\begin{thm}
  \label{thm:cqq:1way:coding}
  For every cqq--state $\rho$,
  $$K_\rightarrow(\rho) \geq I(X;B)-I(X;E).$$
\end{thm}
\begin{beweis}
  The idea is as follows: the state
  $$\rho^{AB}=\sum_x P(x)\ketbra{x}^A\otimes\rho_x^B$$
  contains the description of a cq--channel with channel states $\rho_x$.
  We will cover ``evenly'' all typical type classes of block length
  $n$ by channel codes ${\cal C}_\ell$ to transmit $\approx n I(X;B)$ bits, most of which are
  ``good'' in the sense that they have small error probability. All of them are of
  the kind that the state of $E$, when taking the average over the last
  $\approx n I(X;E)$ bits of the input, is almost a constant operator, $\sigma_\ell$,
  independent of the leading bits.
  \par
  The key distillation scheme works then as follows:
  on observing $x^n$, which is typical with high probability, Alice announces the type
  of it and a random $\ell$, such that $x^n$ is a codeword of the code ${\cal C}_\ell$, to Bob.
  He is able to decode it with high probability (because the code will be good with
  high probability), and they take the leading
  $$\approx n(I(X;B)-I(X;E))\text{ bits}$$
  of the message as the key. This is uniformly distributed because the code is
  entirely within one type class. Eve knows almost nothing about the key since
  she only has a state very close to $\sigma_\ell$, independent of the key.
  \begin{figure}[ht]
    \label{fig:babushka}
    \includegraphics[width=8.5cm]{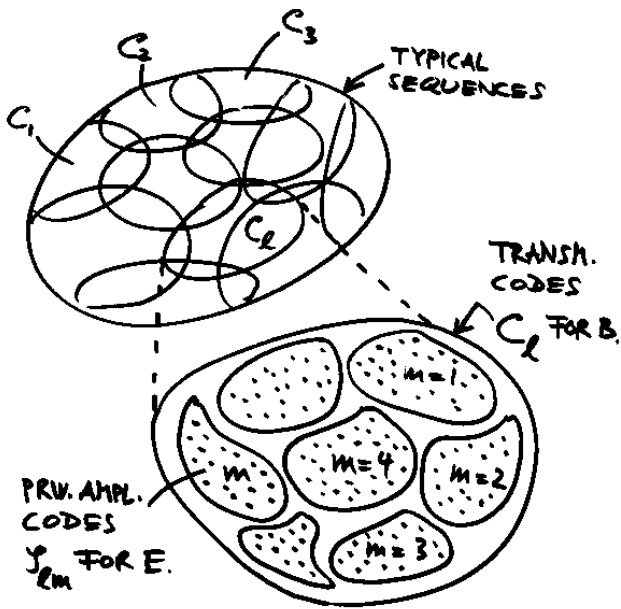}
    \caption{A schematic view of the anatomy of the code:
      the typical sequences are covered by sets
      ${\cal C}_\ell$, which are good transmission codes for $B$. A magnified
      view of one ${\cal C}_\ell$ (to the lower right)
      reveals its inner structure: it is composed
      of ${\cal S}_{\ell m}$, which are good privacy
      amplification codes against $E$.}
  \end{figure}
  \par
  In precise detail: let $Q$ be an $n$--type (Ultimately we will only be interested
  in typical $Q$, i.e. $\|P-Q\|_1 \leq \delta$.)
  Consider random variables $U^{(\ell ms)}$, independent identically distributed
  (i.i.d.) according to the uniform distribution on the type
  class ${\cal T}_Q^n$ (see appendix~\ref{app:types}),
  $\ell=1,\ldots,L$, $m=1,\ldots,M$, $s=1,\ldots,S$.
  Let
  $$\sigma(Q) := \frac{1}{|{\cal T}_Q^n|}\sum_{x^n\in{\cal T}_Q^n} \rho^E_{x^n}
               = \E \rho^E_{U^{(\ell ms)}}.$$
  We are interested in the probability of various random events (for $0<\epsilon<1/2$):
  \par
  {\bf $\epsilon$--Evenness:} for all $x^n\in{\cal T}_Q^n$,
  $$(1-\epsilon)\frac{LMS}{|{\cal T}_Q^n|}
        \leq \sum_{\ell ms} 1_{U^{(\ell ms)}}(x^n) \leq (1+\epsilon)\frac{LMS}{|{\cal T}_Q^n|},$$
  with the indicator functions $1_{U^{(\ell ms)}}$ on ${\cal T}_Q^n$.
  \par
  {\bf $\epsilon$-Secrecy:} for all $\ell,m$, the average of $\rho^E_{x^n}$ over
  ${\cal S}_{\ell m} = \bigl\{ U^{(\ell ms)}:s=1,\ldots,S \bigr\}$
  is close to $\sigma(Q)$:
  $$\left\| \frac{1}{S}\sum_s \rho^E_{U^{(\ell ms)}} - \sigma(Q) \right\|_1 \leq \epsilon.$$
  \par
  {\bf Codes ${\cal C}_\ell$ are $\epsilon$--good:}
  define the code ${\cal C}_\ell$ as the collection of
  codewords $\bigl(U^{(\ell ms)}\bigr)_{m,s}$. We call it \emph{$\epsilon$--good}
  if there exists a POVM $(D^{(\ell)}_{ms})_{m,s}$ such that
  $$\frac{1}{MS}\sum_{ms} \tr\bigl( \rho^B_{U^{(\ell ms)}}D^{(\ell)}_{ms} \bigr) \geq 1-\epsilon.$$
  \par
  Using the Chernoff bound for the indicator functions $1_{U^{(\ell ms)}}$
  evaluated at all points in ${\cal T}_Q^n$ (lemma~\ref{lemma:op:chernoff} and following
  remarks), we obtain
  \begin{equation}
    \label{eq:evenness}
    \Pr\{ \epsilon\text{-evenness} \}
          \geq 1-|{\cal X}|^n\exp\left(\! -LMS\frac{\epsilon^2}{2\ln 2 |{\cal T}_Q^n|} \!\right)\!.
  \end{equation}
  Proposition~\ref{prop:covering} gives us (observing $MS\leq |{\cal X}|^n$), for
  every $\delta>0$ and sufficiently large $n$,
  \begin{equation}
    \label{eq:secrecy}
    \Pr\{ \epsilon\text{-secrecy} \}
          \geq 1-2d^n |{\cal X}|^n\exp\left( -S\iota^n\frac{\epsilon}{288\ln 2} \right),
  \end{equation}
  with $\log\iota = -I(Q;\rho^E)-\delta$.
  \par
  Finally, proposition~\ref{prop:HSW:type} yields
  for every $\delta>0$ and if $MS\leq \exp\bigl( n(I(Q;\rho^B)-\delta) \bigr)$
  ($n$ sufficiently large),
  \begin{equation}
    \label{eq:good-code}
    \forall \ell\quad \Pr\{ {\cal C}_\ell\ \epsilon\text{-good} \} \geq 1-\epsilon.
  \end{equation}
  Since the individual events in this equation are independent, another application
  of the Chernoff bound (to the indiator function of ``$\epsilon$--goodness'') gives,
  \begin{equation}\begin{split}
    \label{eq:most:good-codes}
    \Pr&\{ \text{A fraction }1-2\epsilon\text{ of the }
          {\cal C}_\ell\text{ is }\epsilon\text{-good} \}       \\
       &\phantom{=============:}
             \geq 1-\exp\left( -L\frac{\epsilon^2}{4\ln 2} \right).
  \end{split}\end{equation}
  Thus, if we pick
  \begin{align*}
    S &= \exp\bigl[ n\bigl(I(Q;\rho^E)+2\delta\bigr) \bigr], \\
    M &= \exp\bigl[ n\bigl(I(Q;\rho^B)-I(Q;\rho^E)-3\delta\bigr) \bigr], \\
    L &= \exp\bigl[ n\bigl(H(Q)-I(Q;\rho^B)+2\delta\bigr) \bigr],
  \end{align*}
  and observing $|{\cal T}_Q^n|\leq \exp(nH(Q))$,
  the right hand sides of eqs.~(\ref{eq:evenness}), (\ref{eq:secrecy})
  and (\ref{eq:most:good-codes}) converge to $1$ as $n\rightarrow\infty$,
  and hence by the union bound alse the conjunction of these three events
  approaches unit probability asymptotically.
  \par
  Thus, for sufficiently large $n$, \emph{there exist} codewords
  $u^{(\ell ms)}\in{\cal T}_Q^n$ which together have the property of
  $\epsilon$--evenness, $\epsilon$--secrecy and that a fraction of
  at least $1-2\epsilon$ of the ${\cal C}_\ell=(u^{(\ell ms)})_{m,s}$ is
  $\epsilon$--good.
  Clearly, we can construct such code sets for all types $Q$, of which there
  are at most $(n+1)^{|{\cal X}|}$ many.
  \par
  Now, the protocol works as follows: on observing $x^n$ from the source,
  Alice determines its type $Q$ and sends it to Bob.
  If $x^n$ is not typical, i.e. if $\|P-Q\|_1 > \delta$, the protocol aborts here.
  Otherwise she selects a random
  $\ell$ such that $x^n$ is a codeword of ${\cal C}_\ell$, as well as
  random $m,s$ such that $u^{(\ell ms)}=x^n$. (The latter choice of course
  is unique most of the time: if ${\cal C}_\ell$ is a good code, only a fraction
  of at most $\epsilon$ of the codewords have a collision.)
  She informs Bob also of $\ell$; if ${\cal C}_l$ is not $\epsilon$--good
  the protocol aborts.
  \par
  Note that by the $\epsilon$--evenness of the codewords, the state of $ABE$
  conditional on $Q$ and $\ell$ is
  \begin{equation}
    \label{eq:key:almost}
    \frac{1}{MS}\sum_{m,s} (1\pm\epsilon)\ketbra{ms}^A\otimes\rho^{BE}_{u^{(\ell ms)}}.
  \end{equation}
  (By way of notation, ``$1\pm\epsilon$'' stands for any number in the
  interval $[1-\epsilon;1+\epsilon]$.)
  Now, since ${\cal C}_\ell$ is a good code, Bob can apply the decoding POVM
  $D^{(\ell)}$ to his part of the system, and transform the state in
  eq.~(\ref{eq:key:almost}) into a state $\theta$ with the property
  $$\frac{1}{2}\left\| \theta
                      - \frac{1}{MS}\sum_{m,s}\ketbra{ms}^A\otimes
                                             \ketbra{ms}^B\otimes
                                             \rho^E_{u^{(\ell ms)}} \right\|_1 \leq 2\epsilon.$$
  Both Alice and Bob measure $m$ and end up with a perfectly
  uniformly distributed key of length
  \begin{equation*}\begin{split}
    n&\bigl(I(Q;\rho^B)-I(Q;\rho^E)-3\delta\bigr)   \\
     &\phantom{======}
            \geq n\bigl(I(P;\rho^B)-I(P;\rho^E)-3\delta-\delta'\bigr),
  \end{split}\end{equation*}
  with probability $1-3\epsilon$, where
  $$\delta' = 2\delta\log (d_A d_B d_E)+2\tau(\delta),$$
  with the dimensions $d_B$ and $d_E$ of Bob's and Eve's local
  system, respectively.
  (Recall that $Q$ is typical, and use the Fannes inequality, stated in
  appendix~\ref{app:facts} as lemma~\ref{lemma:Fannes}.)
  By the above property of $\theta$, Alice and Bob disagree with probability $\leq\epsilon$.
  \par
  Finally, thanks to the $\epsilon$--secrecy, for all $\ell$ and $m$,
  $$\left\| \frac{1}{S}\sum_s \rho^E_{u^{(\ell ms)}} -\sigma(Q) \right\|_1 \leq \epsilon,$$
  so Eve's state after the protocol (including her knowledge of $Q$ ond $\ell$)
  is almost constant, whatever the value of $m$.
\end{beweis}
\begin{rem}
  \label{rem:secret-cc}
  The communication cost of the protocol described in the above proof
  is asymptotically
  $$H(X)-I(X;B)=H(A|B)$$
  bits of forward communication (per copy of the state): the information
  which code ${\cal C}_\ell$ to apply from Alice to Bob.
\end{rem}
\par
Here are the facts we use in the proof:
\par
\begin{lemma}[``Operator Chernoff bound''~\cite{Ahlswede:Winter}]
  \label{lemma:op:chernoff}
  Let $X_1,\ldots,X_M$ be i.i.d.~random variables taking values in the operators
  ${\cal B}({\cal H})$ on the $D$--dimensional Hilbert space ${\cal H}$,
  $0\leq X_j\leq \1$, with $A=\E X_j\geq\alpha\1$, and let $0<\eta<1/2$. Then
  \begin{equation*}\begin{split}
    \Pr &\left\{ \frac{1}{M}\sum_{j=1}^M X_j \not\in [(1-\eta)A;(1+\eta)A] \right\}        \\
        &\phantom{===============} \leq 2D \exp\left( -M\frac{\alpha\eta^2}{2\ln 2} \right),
  \end{split}\end{equation*}
  where $[A;B]=\{X: A\leq X\leq B\}$ is an interval in the operator order.
  \qed
\end{lemma}
\par
Note that for the case $D=1$ this reduces to the classical Chernoff bound
for bounded real random variables~\cite{chernoff}. Also the case of finite vectors
of bounded real random variables is included by considering the matrices
with vector entries on the diagonal and zero elsewhere. It is essential in the
proof of the following result.
\par
\begin{prop}
  \label{prop:covering}
  For a cq--channel $W:\fset{X}\longrightarrow {\cal S}({\cal H})$ and a type
  $P$, let $U^{(j)}$ be i.i.d.~according to the uniform distribution on the
  type class ${\cal T}_P^n$, $j=1,\ldots,M$. Define the state
  $$\sigma(P) = \frac{1}{|{\cal T}_P^n|}\sum_{x^n\in{\cal T}_P^n} W^n_{x^n}=\E W^n_{U^{(j)}}.$$
  Then for every $\epsilon,\delta>0$, and sufficiently large $n$,
  \begin{equation*}\begin{split}
    \Pr &\left\{ \left\| \frac{1}{M}\sum_{j=1}^M W^n_{U^{(j)}} - \sigma(P) \right\|_1
                                                               \geq \epsilon \right\}         \\
        &\phantom{============} \leq 2d^n\exp\left( -M\iota^n\frac{\epsilon}{288\ln 2} \right),
  \end{split}\end{equation*}
  with $\log\iota = -I(P;W)-\delta$.
\end{prop}
\begin{beweis}
  The proof is very close to that of the compression theorem for
  POVMs~\cite{winterisation}.
  We reproduce a version of the argument in appendix~\ref{app:proofs}.
\end{beweis}
\par
\begin{prop}[HSW theorem]
  \label{prop:HSW:type}
  Consider a cq--channel $W:\fset{X}\longrightarrow {\cal S}({\cal H})$ and a type
  $P$, and let $U^{(i)}$ be i.i.d.~according to the uniform distribution on the
  type class ${\cal T}_P^n$, $i=1,\ldots,N$. Then for every $\epsilon,\delta>0$
  and sufficiently large $n$, if $\log N\leq n\bigl(I(P;W)-\delta\bigr)$,
  $$\Pr\left\{ {\cal C}=(U^{(i)})_{i=1}^N \text{ is }\epsilon{-good} \right\}
                                                               \geq 1-\epsilon.$$
  Here we call a collection of codewords \emph{$\epsilon$--good} if there exists a POVM
  $(D_i)_{i=1}^N$ on ${\cal H}^{\otimes n}$ such that
  $$\frac{1}{N}\sum_{i=1}^N \tr\bigl( W^n_{U^{(i)}} D_i \bigr) \geq 1-\epsilon.$$
\end{prop}
\begin{beweis}
  This really is only a slight modification of the Holevo--Schumacher--Westmoreland
  argument~\cite{Holevo:coding,SW:coding}: we give the proof in appendix~\ref{app:proofs}.
\end{beweis}
\par\medskip
This coding theorem puts us in the position to prove the
following formula for the one--way secret key distillation capacity
of a cqq--state:
\par
For conditional distributions $Q(u|x)$ and $R(t|u)$ define the states
\begin{thm}
  \label{thm:cqq:1way-key}
  For very cqq--state $\rho$,
  $$K_{\rightarrow}(\rho) = \lim_{n\rightarrow\infty} \frac{1}{n}K^{(1)}\bigl(\rho^{\otimes n}\bigr),$$
  with
  $$K^{(1)}(\rho) = \max_{T|U|X} \bigl[ I(U;B|T)-I(U;E|T) \bigr],$$
  where the maximisation runs over all random variables $U$ depending on $X$
  and $T$ depending on $U$, i.e.~there are channels $Q$ and $R$ such that
  $U=Q(X)$ and $T=R(U)$, and the above formula refers to the state
  \begin{equation*}\begin{split}
    \omega^{TUABE} &= \sum_{t,u,x} R(t|u)Q(u|x)P(x) \\
                   &\phantom{=====}
                      \ketbra{t}^T\otimes\ketbra{u}^U\otimes\ketbra{x}^A\otimes\rho_x^{BE}.
  \end{split}\end{equation*}
  The ranges of $U$ and $T$ may be taken to have cardinalities
  $|T|\leq |{\cal X}|$ and $|U| \leq |{\cal X}|^2$, and furthermore $T$
  can be taken a (deterministic) function of $U$.
\end{thm}
\begin{beweis}
  Let us begin with the converse part, i.e.~the inequality ``$\leq$'':
  Consider an $(n,\epsilon)$--protocol with rate $R$; then
  by its definition, and using standard information inequalities
  and the Fannes inequality lemma~\ref{lemma:Fannes}
  \begin{equation*}\begin{split}
    nR &\leq H(K) + n\bigl(\tau(\epsilon)+\epsilon R\bigr)                       \\
       &\leq I(K:K'\Lambda) + n\bigl(2\tau(\epsilon)+\epsilon R+\epsilon F\bigr) \\
       &\leq I(K;B\Lambda)  + n\bigl(2\tau(\epsilon)+\epsilon R+\epsilon F\bigr) \\
       &\leq I(K;B\Lambda) - I(K;E\Lambda) \\
       &\phantom{===} 
              + n\bigl(3\tau(\epsilon)+\epsilon R+2\epsilon F+\epsilon \log d_E\bigr) \\
       &=    I(K;B|\Lambda) - I(K;E|\Lambda) + n\delta
  \end{split}\end{equation*}
  Letting $U=(K,\Lambda)$ and $T=\Lambda$ we obtain
  $$R \leq \frac{1}{n}K^{(1)}(\rho) + \delta,$$
  with arbitrarily small $\delta$ as $n\rightarrow\infty$.
  \par
  The proof of the properties of $U$ and $T$ is given in appendix~\ref{app:proofs}.
  \par
  Now we come to the proof of the direct part, i.e. the inequality ``$\geq$'':
  it is clearly sufficient to show that, for given $U$ and $T$,
  the rate $R=I(U;B|T)-I(U;E|T)$ is achievable.
  To this end, consider a protocol, where Alice generates $U$ and $T$ for
  each copy of the state i.i.d., and broadcasts $T$: this leaves Alice, Bob
  and Eve in $n$ copies of
  $$\widetilde\rho = \sum_{t,u,x} R(t|u)Q(u|x)P(x)\ketbra{u}^A\otimes
                                                 \rho^{BE}_x\otimes
                                                 \ketbra{t}^{B'}\!\otimes\ketbra{t}^{E'}\!\!.$$
  Observing
  $$R = I(U;BB')-I(U;EE'),$$
  we can invoke theorem~\ref{thm:cqq:1way:coding}, and are done.
\end{beweis}
\par
\begin{rem}
  \label{rem:single-letter}
  Comparing this with the classical analogue in~\cite{AC:1}, it is a slight
  disappointment to see that here we don't get a single--letter formula.
  The reader may want to verify that the technique used there to single--letterise
  the upper bound does not work here, as it introduces conditioning on
  quantum registers, while our $T$ has to be classical.
\end{rem}
\par\medskip
One can clearly also use a general three--party state $\rho^{ABE}$ to
generate secret key between Alice and Bob: a particular strategy certainly
is for Alice to perform a quantum measurement described by the
POVM $Q=(Q_x)_{x\in{\cal X}}$, which leads to the state
$$\widetilde\rho^{A'BE}
      = \sum_x \ketbra{x}^{A'}\otimes\tr_A\bigl( \rho^{ABE}(Q_x\otimes\1^{BE}) \bigr).$$
Then, starting from many copies of the original state $\rho$, we now have
many copies of $\widetilde\rho$, and theorem~\ref{thm:cqq:1way-key} can be applied.
Because we can absorb the channel $U|X$ into the POVM, we obtain the
direct part (``$\geq$'') in the following statement:
\begin{thm}
  \label{thm:qqq:1way-key}
  For every state $\rho^{ABE}$,
  $$K_\rightarrow(\rho) = \lim_{n\rightarrow\infty} K^{(1)}\bigl(\rho^{\otimes n}\bigr),$$
  with
  $$K^{(1)}(\rho) = \max_{Q,T|X} I(X;B|T)-I(X;E|T),$$
  where the maximisation is over all POVMs $Q=(Q_x)_{x\in{\cal X}}$ and
  channels $R$ such that $T=R(X)$,
  and the information quantities refer to the state
  \begin{equation*}\begin{split}
    \omega^{TA'BE} &= \sum_{t,x} R(t|x)P(x)                                \\
                   &\phantom{===}
                                 \ketbra{t}^T\otimes\ketbra{x}^{A'}\otimes         
                                 \tr_A\bigl( \rho^{ABE}(Q_x\otimes\1^{BE}) \bigr).
  \end{split}\end{equation*}
  The range of the measurement $Q$ and the random variable $T$ may assumed
  to be bounded as follows:
  $|T|\leq d_A^2$ and $|{\cal X}|\leq d_A^4$, and furthermore $T$ can be
  taken a (deterministic) function of $X$.
\end{thm}
\begin{beweis}
  After our remarks preceding the statement of the theorem, we have
  only the converse to prove. This will look very similar to the converse
  of theorem~\ref{thm:cqq:1way-key}.
  Even though we haven't so far defined what a key distillation protocol is
  in the present context,
  we can easily do that now (and check that the procedure above
  is of this type): it consists of a measurement POVM $Q=(Q_{\ell m})_{\ell,m=1}^{L,M}$
  for Alice and the POVMs $D^{(\ell)}$ for Bob, with the same conditions
  (1)--(3) as in the first paragraphs of this section, where as before
  we assume a rate bound on the public discussion: $L\leq 2^{nF}$.
  This obviously generalises the definition we gave for cqq--states.
  \par
  Consider an $(n,\epsilon)$--protocol with rate $R$;
  using standard information inequalities
  and the Fannes inequality lemma~\ref{lemma:Fannes} we can estimate
  as follows:
  \begin{equation*}\begin{split}
    nR &\leq H(K) + n\bigl(\tau(\epsilon)+\epsilon R\bigr)                       \\
       &\leq I(K:K'\Lambda) + n\bigl(2\tau(\epsilon)+\epsilon R+\epsilon F\bigr) \\
       &\leq I(K;B\Lambda)  + n\bigl(2\tau(\epsilon)+\epsilon R+\epsilon F\bigr) \\
       &\leq I(K;B\Lambda) - I(K;E\Lambda) \\
       &\phantom{===} 
              + n\bigl(3\tau(\epsilon)+\epsilon R+2\epsilon F+\epsilon \log d_E\bigr) \\
       &=    I(K;B|\Lambda) - I(K;E|\Lambda) + n\delta
  \end{split}\end{equation*}
  The measurement $Q$ and $T(\ell,m)=\ell$ are permissible
  in the definition of $K^{(1)}$, hence we obtain
  $$R \leq \frac{1}{n}K^{(1)}(\rho) + \delta,$$
  with arbitrarily small $\delta$ as $n\rightarrow\infty$.
  \par
  It remains to prove the bounds on the range of $X$ and $T$
  for which we imitate the proof of the corresponding statement in
  theorem~\ref{thm:cqq:1way-key}: the full argument is given
  in appendix~\ref{app:proofs}.
\end{beweis}
\par
\begin{rem}
  \label{rem:worstcase}
  Clearly the worst case for Alice and Bob is when Eve holds the system $E$
  of a purification $\ket{\psi^{ABE}}$ of $\rho^{AB}$, because clearly every other
  extension $\rho^{ABE}$ of $\rho^{AB}$ can be obtained from the purification
  by a quantum operation acting on $E$.
  \par
  Our result (at least in principle) characterises those bipartite states $\rho^{AB}$
  for which one--way key distillation is possible at positive rate. We have to leave open the
  question of characterising the states for which positive rates can be obtained by
  general two--way public discussion (compare the classical
  case~\cite{maurer,AC:1,maurer:wolf,renner:wolf}!).
  \par
  Note that the classical analogue of the ``worst case'' is total knowledge of Eve about
  both Alice's and Bob's random variables --- which makes key distillation totally
  impossible. For quantum states thus, it must be some ``non--classical'' correlation
  which makes positive rates possible; it is tempting to speculate that a manifestation
  of entanglement is behind this effect.
  \par
  We do not fully resolve this issue in the present paper; nevertheless, in a similar 
  vein, we show  in the following section that if $\rho^{AB}$
  allows one-way distillation of EPR pairs at positive rates, 
  then our cryptographic techniques
  give a construction of an entanglement distillation protocol by a modification of
  key distillation protocols of a particular form.
\end{rem}

\section{One--way entanglement distillation}
\label{sec:oneway-ent}
Consider an arbitrary state $\rho^{AB}$ between Alice and Bob. In~\cite{BDSW}
the task of distilling EPR pairs at optimal rate from many copies of $\rho$,
via local operations and classical communication (LOCC), was introduced.
\par
A \emph{one--way entanglement distillation protocol} consists of
\begin{itemize}
  \item A quantum instrument ${\bf T}=(T_\ell)_{\ell=1}^L$ for Alice.
    (An \emph{instrument}~\cite{davies:lewis} is a quantum operation with both
    classical and quantum outputs --- it is modelled in general as a cp--map
    valued measure; for our purposes it is a finite collection of cp--maps
    which sum to a cptp map.)
  \item For each $\ell$ a quantum operation $R_\ell$ for Bob.
\end{itemize}
\par
We call it an \emph{$(n,\epsilon)$--protocol} if it acts on $n$ copies
of the state $\rho$ and produces a maximally entangled state
$$\ket{\Phi_M} = \frac{1}{\sqrt{M}}\sum_{m=1}^M \ket{m}^A\otimes\ket{m}^B$$
up to fidelity $1-\epsilon$:
$$F\left( \Phi_M,\sum_{\ell=1}^L (T_\ell\otimes R_\ell)\bigl(\rho^{\otimes n}\bigr) \right)
                                                                             \geq 1-\epsilon.$$
Note that we may assume without loss of generality that $T_\ell$ and $R_\ell$
output states supported on the reduced states of $\Phi_M$ on Alice's and Bob's
system, respectively: otherwise we could improve the fidelity.
\par
A number $R$ is an \emph{achievable rate} if there exist, for every $n$,
$(n,\epsilon)$-protocol, with $\epsilon\rightarrow 0$ and
$\frac{1}{n}\log M\rightarrow R$ as $n\rightarrow\infty$. Finally,
$$D_\rightarrow(\rho) := \sup\{ R:R\text{ achievable} \}$$
is the \emph{one--way (or forward) entanglement capacity of $\rho$}.
\par
In~\cite{BDSW} the case of Bell--diagonal two--qubit states,
$$\rho = p_{00}\Phi^+ + p_{01}\Phi^- + p_{10}\Psi^+ + p_{11}\Psi^-,$$
was considered and proved that $D_\rightarrow(\rho)\geq 1-H(\{p\})$,
by a method called ``hashing protocol'' (this was generalised
recently to higher dimensions in~\cite{vollbrecht:wolf}).
Concerning lower bounds not much more is
known, but there are numerous works dealing with upper bounds on the distillable
entanglement: the \emph{entanglement of formation} $E_F(\rho)$~\cite{BDSW}, the
\emph{relative entropy of entanglement} $E_{\rm re}(\rho)$~\cite{vedral:etal},
the \emph{Rains bound} $R(\rho)$~\cite{rains}, and the recently introduced
\emph{squashed entanglement} $E_{\rm sq}(\rho)$~\cite{upperbound}.
\par
To connect to the cryptographic setting discussed so far, construct a purification
$\ket{\psi}^{ABE}$ of $\rho$, of which we are particularly interested in its
Schmidt form
$$\ket{\psi}^{ABE} = \sum_x \sqrt{P(x)}\ket{x}^A\otimes\ket{\psi_x}^{BE}.$$
\par
Consider the following special strategy for a one--way secret key distillation
protocol, on $n$ copies of the state: Alice measures $x^n$ (i.e. the above Schmidt
basis), and applies the secret key distillation protocol from
theorem~\ref{thm:cqq:1way:coding}:
it is easy to evaluate the key rate:
\begin{equation*}\begin{split}
  I(P;\psi^B)-I(P;\psi^E) &= H(B)-H(E) \\
                          &= H(B)-H(AB).
\end{split}\end{equation*}
\par
By letting Alice and Bob execute this protocol ``coherently'', we can prove:
\begin{thm}[Hashing inequality]
  \label{thm:hashing}
  $$D_{\rightarrow}(\rho) \geq H(\rho^B)-H(\rho^{AB}).$$
  The right hand side here equals the negative conditional von Neumann
  entropy, $-H(A|B)$, a quantity known as
  \emph{coherent information}~\cite{coherent}, which we denote
  (acknowledging is directionality) $I_c(A\,\rangle B)$. If the state
  this is referring to is not apparent from the context, we add it
  in subscript: $I_c(A\,\rangle B)_\rho$.
\end{thm}
\begin{beweis}
  Recall the structure of the protocol in the proof of
  theorem~\ref{thm:cqq:1way:coding}:
  for each typical type $Q$ we have a collection of codewords $u^{(\ell ms)}$,
  $\ell=1,\ldots,L$, $m=1,\ldots,M$ and $s=1,\ldots,S$ from
  ${\cal T}_Q^n$ satisfying $\epsilon$--evenness, $\epsilon$--secrecy,
  and a fraction of at least $1-2\epsilon$ of the codes
  ${\cal C}_\ell=\bigl( u^{(\ell ms)} \bigr)_{m,s}$ are $\epsilon$-good.
  \par
  The first step of the protocol is that
  Alice measures the type $Q$ non--destructively and informes Bob about
  the result. The protocol aborts if $Q$ is not typical, i.e. if
  $\| P-Q \|_1 > \delta$. This leaves the post--measurement state
  $$\sqrt{\frac{1}{|{\cal T}_Q^n|}}\sum_{x^n\in{\cal T}_Q^n}
                                                  \ket{x^n}^A\otimes\ket{\psi_{x^n}}^{BE}.$$
  \par
  Define now a quantum operation for Alice, with Kraus elements
  \begin{align*}
    C_{\ell}    &= \sqrt{\frac{1}{1+\epsilon}\frac{|{\cal T}_Q^n|}{LMS}}
                                       \sum_{m,s} \ket{ms}\!\bra{u^{(\ell ms)}}, \\
    C_\emptyset &= \sqrt{\1-\sum_{Q,\ell} C_{\ell}^\dagger C_{\ell}},
  \end{align*}
  which we interpret as an instrument with outcomes $\ell$ and
  $\emptyset$~\cite{davies:lewis}: $T_\ell(\sigma)=C_\ell\sigma C_\ell^\dagger $.
  This outcome is communicated to Bob.
  That these are really permissible Kraus operators  we obtain
  from the $\epsilon$--evenness condition.
  \par
  The outcome $\emptyset$ (resulting in abortion of the protocol)
  is observed with probability at most $\epsilon$
  (if $n$ is large enough). The other outcomes $\ell$ all occur with the probability
  $$\gamma(Q) = P^{\otimes n}({\cal T}_Q^n)\frac{1}{1+\epsilon}\frac{1}{L},$$
  in which case the output state of the instrument is
  \begin{equation}
    \label{eq:ent:almost}
    \sqrt{\frac{1}{MS}} \sum_{ms} \ket{ms}^A\otimes\ket{\psi_{u^{(\ell ms)}}}^{BE}.
  \end{equation}
  (The absence of the $1\pm\epsilon$ factors when compared to the analogous
  eq.~(\ref{eq:key:almost}) in the proof of theorem~\ref{thm:cqq:1way:coding})
  is due to our having introduced the error event $\emptyset$.)
  \par
  Now, just as in the proof of theorem~\ref{thm:cqq:1way:coding}, Bob decodes
  $m$ and $s$, at least if ${\cal C}_\ell$ is $\epsilon$--good (which
  fails to happen with probability only $2\epsilon$). But he does it coherently,
  by introducing an ancilla system $B'$ in a standard state $\ket{0}$ and applying
  a unitary to extract $ms$ into $B$, leaving in $B'$ whatever is necessary to make
  the map unitary.
  This transforms the state in eq.~(\ref{eq:ent:almost}) into a state
  \begin{equation*}\begin{split}
    &\ket{\vartheta}^{ABB'E}
       = \sqrt{\frac{1}{MS}}\sum_{ms} \ket{ms}^A\otimes \\
    &\phantom{==:}
                 \otimes\!\left( \sqrt{1-e_{ms}}\ket{ms}^B\ket{\varphi^{\rm OK}_{\ell ms}}^{B'E}
                                \!\!+\!\sqrt{e_{ms}}  \ket{\varphi^{\rm bad}_{\ell ms}}^{BB'E} \right)\!\!,
  \end{split}\end{equation*}
  where $e_{ms}$ is the probability of the code incorrectly identifying $ms$,
  and $\ket{\varphi^{\rm bad}_{\ell ms}}$ is orthogonal to $\ket{ms}\ket{\varphi^{\rm OK}_{\ell ms}}$.
  Now, because the code is $\epsilon$--good,
  \begin{equation*}\begin{split}
    F&\!\left( \ket{\vartheta},
            \sqrt{\frac{1}{MS}}\sum_{ms} \ket{ms}^A\otimes
                                         \ket{ms}^B\ket{\varphi^{\rm OK}_{\ell ms}}^{B'E} \right) \\
     &\phantom{====================:}
                                                                       \geq (1-\sqrt{\epsilon})^2 \\
     &\phantom{====================:}
                                                                       \geq 1-2\sqrt{\epsilon},
  \end{split}\end{equation*}
  where we have used the Markov inequality: at most a fraction of $\sqrt{\epsilon}$
  of the $e_{ms}$ can be larger than $\sqrt{\epsilon}$.
  Since the decoding only affects Bob's registers, but certainly not $E$, we have
  \begin{equation*}
    (1-e_{ms})\left(\varphi^{\rm OK}_{\ell ms}\right)^E
      + e_{ms}\left(\varphi^{\rm bad}_{\ell ms}\right)^E
                                                          = \psi_{u^{(\ell ms)}}^E,
  \end{equation*}
  and hence we can assume that
  \begin{equation}
    \label{eq:eveisout}
    \bra{\varphi^{\rm OK}_{\ell ms}} \psi_{u^{(\ell ms)}}\rangle \geq \sqrt{1-e_{ms}}.
  \end{equation}
  This implies
  \begin{equation}\begin{split}
    \label{eq:ent:fidelity}
    F&\!\left( \ket{\vartheta},
            \sqrt{\frac{1}{MS}}\sum_{ms} \ket{ms}^A\otimes
                                         \ket{ms}^B\ket{\psi_{u^{(\ell ms)}}}^{B'E} \right) \\
     &\phantom{====================:}
                                                                   \geq 1-3\sqrt{\epsilon}.
  \end{split}\end{equation}
  At this point, Alice and Bob almost have their maximal
  entanglement of the $m$--variable. All that remains is to be
  done is to disentangle Eve:
  \par
  To begin, Alice measures the $s$--component of her register
  in the Fourier--transformed basis:
  $$\left( \ket{\hat{t}}=\sqrt{\frac{1}{S}}
                             \sum_{s=1}^S e^{2\pi i st/S}\ket{s}:t=1,\ldots,S \right),$$
  and tells Bob the result $t$, who applies the phase shift
  $$\sum_{s=1}^S e^{2\pi i st/S}\ketbra{s}$$
  to the $s$--component of his register $B$. This
  transforms $\ket{\vartheta}^{ABB'E}$ into a state $\ket{\Theta}^{ABB'E}$ with
  \begin{equation}
    \label{eq:ent:FIDELITY}
    F\!\left( \!\ket{\Theta},
                \sqrt{\frac{1}{MS}}\sum_{m} \ket{m}^A\!\otimes\!
                                            \ket{ms}^B\ket{\varphi^{\rm OK}_{\ell ms}}^{B'E} \right)\!
    \geq 1-3\sqrt{\epsilon},
  \end{equation}
  invoking the non--decrease of the fidelity under quantum operations,
  applied to eq.~(\ref{eq:ent:fidelity}).
  \par
  Absorbing $s$ into the register $B'$,
  the right hand state in the last equation can be rewritten as
  \begin{equation}
    \label{eq:only:s:remains}
    \frac{1}{\sqrt{M}}\sum_m \ket{m}^A\otimes\ket{m}^B\ket{\widetilde\psi_{\ell ms}}^{B'E},
  \end{equation}
  with
  $$\ket{\widetilde\psi_{\ell ms}}^{B'E}
        = \frac{1}{\sqrt{S}}\sum_s \ket{s}^{B'_1}\ket{\hat\psi_{u^{(\ell ms)}}}^{B'_2E}.$$
  The reduced states of Eve of the $\ket{\widetilde\psi_{\ell ms}}^{B'E}$ is
  $$\sigma_{\ell m}=\frac{1}{S}\sum_s \psi^E_{u^{(\ell ms)}},$$
  where we made use of eq.~(\ref{eq:eveisout}), which is, by the
  $\epsilon$--secrecy, at trace distance at most $\epsilon$ 
  from a state we denoted $\sigma(Q)$ in the proof of theorem~\ref{thm:cqq:1way:coding}.
  By lemma~\ref{lemma:norm-fidelity} in appendix~\ref{app:facts},
  $F\bigl( \sigma_{\ell m},\sigma(Q) \bigr) \geq 1-\epsilon$.
  \par
  Choosing a purification $\ket{\zeta}^{B'E}$ of $\sigma(Q)$, this means that there are
  unitaries $U_{\ell m}$ on $B'$ such that
  $$F\left( (U_{\ell m}\otimes\1)\ket{\widetilde\psi_{\ell ms}},\ket{\zeta} \right)
                                                                       \geq 1-\epsilon,$$
  because the mixed state fidelity equals the maximum pure state fidelity
  over all purifications of the states and all purifications are related by
  unitaries on the purifying system~\cite{uhlmann,jozsa}.
  Hence, if Bob applies
  $$U:=\sum_m \ketbra{m}\otimes U_{\ell m}$$
  to his share of the state,
  then the state in eq.~(\ref{eq:only:s:remains}) is transformed into
  a state $\ket{\Xi}^{ABB'E}$ with
  \begin{equation*}
    F\left( \ket{\Xi},\frac{1}{\sqrt{M}}\sum_m \ket{m}^A\otimes
                                               \ket{m}^B\ket{\zeta}^{B'E} \right)
    \geq 1-\epsilon.
  \end{equation*}
  Of course, he actually works on $\ket{\Theta}$, so they end up with
  the state $(\1\otimes U\otimes\1)\ket{\Theta}$, which has fidelity
  $1-3\sqrt{\epsilon}$ to $\ket{\Xi}$, hence with eq.~(\ref{eq:ent:FIDELITY})
  we conclude (by simple geometry) that it has fidelity $\geq 1-12\sqrt{\epsilon}$
  to $\ket{\Phi_M}^{AB}\otimes\ket{\zeta}^{B'E}$.
  \par
  Nontypical $Q$, the event $\emptyset$ or bad code ${\cal C}_\ell$
  happen with total probability at most $4\epsilon$. In the ``good'' case,
  Alice and Bob distill --- up to fidelity
  $1-12\sqrt{\epsilon}$ --- a maximally entangled state of
  log Schmidt rank
  \begin{equation*}\begin{split}
    n&\bigl( I(Q;\psi^B)-I(Q;\psi^E)-3\delta \bigr)                   \\
     &\phantom{======}
          \geq n\bigl( I(P;\psi^B)-I(P;\psi^E)-3\delta-\delta' \bigr) \\
     &\phantom{======}
          =    n\bigl( H(B)-H(E)-3\delta-\delta' \bigr).
  \end{split}\end{equation*}
  with $\delta'$ just as at the end of the proof of theorem~\ref{thm:cqq:1way:coding}.
\end{beweis}
\par
\begin{rem}
  \label{rem:ent-cc}
  The communication cost of the above protocol is asymptotically
  \begin{equation*}\begin{split}
    H(A)-I(X;B)+I(X;E) &= H(A)-H(B)+H(E)  \\
                       &= H(A)+H(E)-H(AE) \\
                       &= I(A:E)
  \end{split}\end{equation*}
  bits of forward classical communication per copy of the state:
  the information which code ${\cal C}_\ell$ to use, plus the
  information from the measurement of the Fourier--transformed
  basis $(\ket{\hat{t}})_t$.
  \par
  Even though at first sight there seems to be little reason
  to believe that our procedure is
  optimal for this resource (consider for example a separable initial state:
  Alice will have mutual information with a purification but clearly the best
  thing is to do nothing), it is amusing to see the quantum mutual
  information show up here.
  \par
  It is in fact possible to show that subject to another optimisation,
  the quantum mutual information between Alice and Eve gives indeed the
  minimum forward communication cost~\cite{D:H:W}.
\end{rem}
\begin{expl}
  \label{expl:bell-states}
  It is interesting to compare our method to the original hashing protocol
  of~\cite{BDSW}, for the case of mixtures of Bell states
  $$\rho = \sum_{i,j=0}^1 \p_{ij}\Phi_{ij},$$
  with the numbering of the Bell states introduced in~\cite{BDSW}:
  $$\Phi_{00}=\Phi^+,\ \Phi_{01}=\Phi^-,\ \Phi_{10}=\Psi^+,\ \Phi_{11}=\Psi^-.$$
  The purification we use in the proof reads
  \begin{equation*}\begin{split}
    \ket{\psi}^{ABE} &= \sum_{i,j=0}^1 \sqrt{p_{ij}}\ket{\Phi_{ij}}^{AB}\otimes\ket{ij}^E \\
                     &= \frac{1}{\sqrt{1}}\bigl( \ket{0}^A\ket{\psi_0}^{BE}
                                                +\ket{1}^A\ket{\psi_1}^{BE} \bigr),
  \end{split}\end{equation*}
  with
  \begin{align*}
    \ket{\psi_0}^{BE} &= \sqrt{p_{00}}\ket{0}^B\ket{00}^E
                           +\sqrt{p_{01}}\ket{0}^B\ket{01}^E  \\
                      &\phantom{===}
                        +\sqrt{p_{10}}\ket{1}^B\ket{10}^E
                           +\sqrt{p_{11}}\ket{1}^B\ket{11}^E, \\
    \ket{\psi_1}^{BE} &= \sqrt{p_{00}}\ket{1}^B\ket{00}^E
                           -\sqrt{p_{01}}\ket{1}^B\ket{01}^E  \\
                      &\phantom{===}
                        +\sqrt{p_{10}}\ket{0}^B\ket{10}^E
                           -\sqrt{p_{11}}\ket{0}^B\ket{11}^E.
  \end{align*}
  Note that this is indeed a Schmidt decomposition.
  First of all, the communication cost of our protocol evaluates
  (using the symmetry between A and B) to
  \begin{equation*}\begin{split}
    I(A:E) &= I(B:E)          \\
           &= H(B)+ H(E) - H(BE) \\
           &= 1   + H(\{p\}) - 1 = H(\{p\}),
  \end{split}\end{equation*}
  which is the same as in~\cite{BDSW}. But the way of the hashing protocol is
  to ``hash'' information about the identity of the state in the Bell ensemble
  into approximately $\approx nH(\{p\})$ of the states, which then are measured
  locally and the results communicated. Our protocol in contrast has two very
  distinct communication parts: there is the ``code information'' (which
  amounts to error correction between Alice and Bob, with built--in
  privacy amplification for Eve's information about the basis state),
  and there is the ``phase information'' from the measurement in the
  Fourier transformed basis. The first amounts to
  $$H(X)-I(X;\psi^B) = H(p_{00}+p_{01},p_{10}+p_{11}),$$
  while the second is
  $$I(X;E) = H(\{p\}) - H(p_{00}+p_{01},p_{10}+p_{11}).$$
\end{expl}
\par\medskip
Our result leads to the general formula for one--way distillable entanglement:
\begin{thm}
  \label{thm:1way-ent}
  For any bipartite state $\rho^{AB}$,
  $$D_{\rightarrow}(\rho) = \lim_{n\rightarrow\infty} \frac{1}{n}D^{(1)}\left(\rho^{\otimes n}\right),$$
  with
  $$D^{(1)}(\rho) := \max_{\bf T} \sum_{\ell=1}^L \lambda_\ell I_c(A\,\rangle B)_{\rho_\ell},$$
  where the maximisation is over quantum instruments ${\bf T}=(T_1,\ldots,T_L)$ on Alice's system,
  $\lambda_\ell=\tr\, T_\ell(\rho^A)$ and $\rho_\ell = \frac{1}{\lambda_\ell}(T_\ell\otimes\id)(\rho)$.
  The range of $\ell$ can be assumed to be bounded, $L\leq d_A^2$, and moreover
  each $T_\ell$ can be assumed to have only one Kraus operator:
  $T_\ell(\sigma)=A_\ell\sigma A_\ell^\dagger $.
\end{thm}
\begin{beweis}
  First, for the direct part, it is sufficient to consider an instrument ${\bf T}$ on
  one copy of the state:
  if Alice performs the instrument ${\bf T}$ on each copy
  and communicates the result to
  Bob, they end up with the new state
  $$\widetilde\rho = \sum_\ell \lambda_\ell\rho_\ell^{AB}\otimes\ketbra{\ell}^{B'}.$$
  Observe that
  $$I_c(A\,\rangle BB')_{\widetilde\rho} = \sum_\ell \lambda_\ell I_c(A\,\rangle B)_{\rho_\ell},$$
  thus application of theorem~\ref{thm:hashing} to $\widetilde\rho$ gives achievability.
  \par
  For the converse, consider any one--way distillation protocol with rate $R$,
  and denote Alice's instrument by ${\bf T}=(T_\ell)_\ell$, Bob's quantum
  operations by $R_\ell$. Write
  \begin{equation*}
    \Omega =  \sum_\ell (T_\ell\otimes R_\ell)(\rho^{\otimes n})
           =: \sum_\ell \lambda_\ell \Omega_\ell.
  \end{equation*}
  Then, using Fannes inequality lemma~\ref{lemma:Fannes}, the convexity
  of the coherent information in the state~\cite{lieb:ruskai} and
  quantum data processing~\cite{coherent},
  \begin{equation*}\begin{split}
    nR &\leq H(\Omega^B) - H(\Omega^{AB}) + 2n\bigl( \tau(\epsilon)+\epsilon R \bigr) \\
       &=    I_c(A\,\rangle B)_\Omega + 2n\bigl( \tau(\epsilon)+\epsilon R \bigr)     \\
       &\leq \sum_\ell \lambda_\ell I_c(A\,\rangle B)_{\Omega_\ell}
                                      + 2n\bigl( \tau(\epsilon)+\epsilon R \bigr)     \\
       &\leq \sum_\ell \lambda_\ell I_c(A\,\rangle B)_{\omega_\ell}
                                      + 2n\bigl( \tau(\epsilon)+\epsilon R \bigr),
  \end{split}\end{equation*}
  where $\omega_\ell = \frac{1}{\lambda_\ell}(T_\ell\otimes\id)(\rho^{\otimes n})$.
  Hence we get
  $$R \leq \frac{1}{n}D^{(1)}(\rho^{\otimes n}) + \delta',$$
  with arbitrarily small $\delta'$ as $n\rightarrow\infty$, and we are done.
  \par
  As for the bound on $L$ and the structure of ${\bf T}$, observe that if
  one $T_\ell$ has more than one Kraus element, one can decompose $T_\ell(\sigma)$
  into a sum of terms $A_{\ell j}\sigma A_{\ell j}$: for the corresponding
  probabilities $\lambda_\ell = \sum_j \lambda_{\ell j}$ and for the post--measurement
  states $\lambda_\ell\rho_\ell = \sum_j \lambda_{\ell j}\rho_{\ell j}$.
  Then by the convexity of $I_c$ in the state~\cite{lieb:ruskai},
  $$\lambda_\ell I_c(A\,\rangle B)_{\rho_\ell}
                \leq \sum_j \lambda_{\ell j}I_c(A\,\rangle B)_{\rho_{\ell j}}.$$
  By the polar decomposition and invariance of $I_c$ under local unitaries
  we may further assume that $A_\ell\geq 0$, i.e.
  $A_\ell = \sqrt{A_\ell^\dagger A_\ell}$; in this form the whole instrument is actually
  described by the POVM $\bigl(A_\ell^2\bigr)_\ell$, and each POVM corresponds
  to an instrument by taking as the $A_\ell$ the square roots of the POVM operators.
  \par
  Now, invoking a theorem of Davies~\cite{davies} (which actually
  is another application of Caratheodory's theorem, lemma~\ref{lemma:caratheodory}),
  any POVM is a convex combination of extremal POVMs, which have at most $d_A^2$
  non--zero elements each, and this convex decomposition clearly carries over to
  the instruments: ${\bf T} = \sum_j \pi_j{\bf T}_j$.
  Since then $\sum_\ell \lambda_\ell I_c(A\,\rangle B)_{\rho_\ell}$
  is the same convex combination of similar such terms for the  instruments
  ${\bf T}_j$, at least one of these gives a higher yield
  $\sum_\ell \lambda_{\ell j} I_c(A\,\rangle B)_{\rho_\ell}$:
  note that the cp--maps of ${\bf T}_j$ are scalar multiples of the $T_\ell$,
  hence the output state of ${\bf T}_j$ with classical result $\ell$
  is $\rho_\ell$.
\end{beweis}

\section{Quantum and entanglement capacities}
\label{sec:cap-and-ent}
Horodecki${}^3$~\cite{HHH} have observed that the hashing inequality implies
information theoretic formulas for a number of quantum capacities and the distillable
entanglement:
\par
In particular, the quantum capacity of a quantum channel, either
unassisted or assisted by forward or two--way communication is given by a formula
involving coherent information (where we indicate the assisting resource
in the subscript):
\begin{thm}
  \label{thm:Q:capacities}
  Let $T:{\cal B}({\cal H}_A)\longrightarrow{\cal B}({\cal H}_B)$ be
  any quantum channel. Then,
  $$Q_\emptyset(T) = Q_\rightarrow(T)
                   = \lim_{n\rightarrow\infty}
                        \frac{1}{n}\max_{\ket{\psi}} I_c(A'\,\rangle B^n)_\omega,$$
  with any pure state $\psi$ on $A'A^n$ and the state
  $$\omega = (\id\otimes T)^{\otimes n}(\psi^{A'A}).$$
  Furthermore,
  $$Q_\leftrightarrow(T) = \lim_{n\rightarrow\infty}
                              \frac{1}{n}\sup_{\ket{\psi,V}} I_c(A'\,\rangle B^n)_\omega,$$
  with any pure state $\psi$ on $A'A^n$, two--way LOCC operation
  $V$ and the state
  $$\omega = V\left[ (\id\otimes T)^{\otimes n}(\psi^{A'A}) \right].$$
\end{thm}
\begin{beweis}
  See~\cite{HHH}. That forward communication does not help was proved
  in~\cite{BKN}, and that the right hand side is an upper bound to $Q_\emptyset$
  was shown in~\cite{coherent,Lloyd:Q}.
  \par
  The idea of achievability is to distill the state $\omega$ and then use
  teleportation --- this involves forward communication but either it is
  free or the whole procedure including the distillation and teleportation
  uses only forward communication, which by~\cite{BKN} can be removed.
  \par
  In~\cite{HHH} a similar formula (involving a coherent
  information $I_c(B\,\rangle A)_\sigma$) was proposed for
  the quantum capacity with classical feedback. However, the proof as indicated
  above does not work in this case: indeed we may use the back--communication
  to help distillation, but teleportation needs a \emph{forward} communication,
  so we end up with a quantum channel code utilising two--way classical
  side communication, which is not known to be reducible
  to just back--communication: in fact, the results of
  Bowen~\cite{bowen} might be taken as indication that for the
  erasure channel the capacities with feedback and with two--way
  side communication are different.
\end{beweis}
\par\medskip
We have already given a formula for the distillable entanglement using
one--way LOCC in theorem~\ref{thm:1way-ent}:
\begin{thm}
  \label{thm:ent:capacities}
  For any state $\rho^{AB}$,
  $$D(\rho) = \lim_{n\rightarrow\infty}
                              \frac{1}{n}\sup_{V} I_c(A'\,\rangle B')_\omega,$$
  with any two--way LOCC operation $V$ and the coherent information
  refers to the state $\omega = V\bigl( \rho^{\otimes n} \bigr)$.
\end{thm}
\begin{beweis}
  For the direct part (``$\geq$'') it is obviously sufficient to consider
  any two--way LOCC operation $V$ on the bipartite system, which applied to
  $\rho$ gives a state $\sigma$. Doing that for $n$ copies of $\rho$,
  application of theorem~\ref{thm:hashing} shows that we can distill
  EPR pairs at rate $I_c(A'\rangle B')_\sigma$ from this.
  \par
  Conversely, let $V_0$ be a two--way LOCC producing a state $\Omega$ with
  $\| \Omega-\Phi_M \|_1\leq \epsilon$, $nR=\log M$. Without loss of generality
  we may assume that $\Omega$ is supported within the tensor product of the
  supports of the reduced states of $\Phi_M$. Thus,
  \begin{equation*}\begin{split}
    nR &\leq H(\Omega^B)-H(\Omega^{AB}) + 3n\bigl( \epsilon R+\tau(\epsilon) \bigr) \\
       &=    I_c(A\,\rangle B)_\Omega   + 3n\bigl( \epsilon R+\tau(\epsilon) \bigr) \\
       &\leq \sup\bigl\{ I_c(A\,\rangle B)_{V(\rho^{\otimes n})} : V \text{ two--way LOCC} \bigr\} \\
       &\phantom{==}                                   + 3n\bigl( \epsilon R+\tau(\epsilon) \bigr),
  \end{split}\end{equation*}
  and we are done.
\end{beweis}
\par\medskip
It was shown furthermore in~\cite{HHH} that for an
ensemble $\{p_i,\rho_i\}$ of bipartite pure states
the hashing inequality implies
\begin{equation*}\begin{split}
  \Delta D &:=    \sum_i p_i D(\rho_i) - D(\rho)                 \\
           &\!\!\leq \Delta I := H(\rho) - \sum_i p_i H(\rho_i).
\end{split}\end{equation*}
This inequality was first exhibited in~\cite{eisert:et-al} for a class
of examples, and conjectured to be true in general. Note that the inequality
is trivially true (using only concavity of the entropy) for the
\emph{loss of coherent information} on the left hand side.
\par\bigskip\noindent
{\bf History and relation to other work:}
\par
The coherent information made its appearance in~\cite{coherent} where its relation
to quantum channel capacity was conjectured and many of its properties proved.
Independently~\cite{Lloyd:Q} proposed this quantity and a heuristic for a proof
which however fell short of a proof. Only recently Hamada~\cite{hamada} succeeded
in giving a lower bound on quantum channel capacity in terms of coherent information
--- still with a crucial restriction to stabiliser codes. It took until~\cite{shor:Q}
for a full proof to be found --- but then quite quickly one of us~\cite{devetak}
discovered a proof based on private information transmission, an idea inspired
by the work of Schumacher and Westmoreland~\cite{SW:private}.
\par
Regarding entanglement distillation, the hashing inequality appears to have been a folk
conjecture from the publication of~\cite{BDSW} on, which however has received much less
attention than the quantum channel coding problem. It was codified as an important
conjecture in~\cite{HHH}.
\par
While completing the writing of the present paper we learned of the work~\cite{scoop},
in which it is shown that the proof by random coding of the channel capacity theorem
can be used to obtin the hashing inequality. It may be interesting to compare
the proofs~\cite{shor:Q,scoop} for the achievability of the coherent information
to ours and~\cite{devetak}. While we, on the face of it, take a detour via
secret key distillation, the final procedure can be argued more direct:
in particular, we don't require the ``double blocking'' which in the
other approaches seem necessary to reduce to a situation in which Alice's
end is in a maximally mixed state. Thus, presumably, our codes achieve rates
approaching the coherent information more quickly, i.e.~for smaller
block length.

\section{Conclusion}
\label{sec:conclusion}
Our findings not only transport an existing classical theory of distilling
secret key from prior correlation (Maurer~\cite{maurer}, Ahlswede and
Csisz\'{a}r~\cite{AC:1}, and follow--up work) to the quantum case, but also
link this subject to entanglement distillation in an operational way:
a coherent implementation of the basic secret key distillation protocol
yielded an entanglement distillation protocol achieving the coherent
information --- this then implies information theoretic formulas for
distillable entanglement and quantum capacities.
\par
We want to draw the reader's attention to several open question
that we have to leave: first of all, are there states for which
$D_\rightarrow(\rho) < K_\rightarrow(\rho)$? Are there maybe even
bound entangled states with positive key rate? A first step might be
to find states such that $D^{(1)}(\rho) < K^{(1)}(\rho)$.
Note that the potential gap between $K_\rightarrow$ and $D_\rightarrow$
comes from the possibility to have more general measurements
at Alice's side than the complete von Neumann measurement in the Schmidt basis
that was our starting point in the proof of theorem~\ref{thm:hashing}
(actually any complete von Neumann measurement would do): namely,
in key distillation, a viable option for Alice is to discard
part of her state (corresponding to using higher rank POVM
elements), but keep that part secret from Eve all the same;
while in entanglement distillation, `Eve'' is everything
except Alice and Bob, so it is as if she would get the parts Alice
decided to toss away.
\par
A second group of open questions:
in general, the optimisations in theorems~\ref{thm:cqq:1way-key},
\ref{thm:qqq:1way-key} and~\ref{thm:1way-ent} are quite nasty,
most so because they involve a limit of many copies of the state.
In the classical theory of secret key distillation, a single--letter
formula for the optimal one--way key rate is proved in~\cite{AC:1},
so there might be hope at least for theorem~\ref{thm:cqq:1way-key}.
In contrast, the optimal rate of two--way protocols or even a procedure
to decide if it is nonzero is still to be found (see the very well--informed
reviews~\cite{michal:e-measures} and~\cite{pawel:ryszard}),
which is why we concentrate on one--way protocols for now. 
It is known that distillability of entanglement may be absent for a single
copy of a state, but could appear for collective operations on several copies
(see again the review~\cite{pawel:ryszard}, sections 6.3 and 7.2
and references therein), so there are only limited possibilities
for making theorem~\ref{thm:1way-ent} into a single--letter formula.
Note in particular that the results of~\cite{dVSS} --- see also
the discussion of Barnum, Nielsen and Schumacher~\cite{coherent}
where the failure of subadditivity for the coherent information is
observed ---
imply that single--letter maximisation of the coherent information
will certainly not achieve the optimum distillability.
It would therefore be good to have at least
an a priori bound on the number $n$ of copies of the state which we have to
consider to have $D^{(1)}(\rho^{\otimes n})$ within, say $\epsilon$,
of the optimal rate. In general, good single--letter lower and upper
bounds~\cite{upperbound} are still wanted!
\par
Finally, we would like to know what the public/classical communication cost
is of distilling secret key and entanglement, respectively, in particular
in the one--way scenario (which at any rate seems to be the one open to
analysis). More generally, if we limit the amount of communication, can we determine
the optimal distillation rates (see \cite{CR} for the communication cost
of \emph{common randomness} distillation)? In the entanglement case this should link up
with initial efforts to understand the communication cost of various
state transformation tasks~\cite{Lo,Harrow:Lo,Hayden:Winter,Ambainis}.
A study concerning the forward
communication cost of entanglement distillation is in perparation~\cite{D:H:W}.

\acknowledgments
We wish to thank Patrick Hayden and Debbie Leung for a discussion at the
conception of this work.
After the present research had been done, we learned from
Micha\l\ and Pawe\l\ Horodecki that they had a different approach to
proving the hashing inequality, which they were able to carry through
after hearing of our result. We want to thank them for showing
us an early draft of their paper~\cite{scoop}.
\par
ID is supported in part by the NSA under US Army Research Office (ARO) grants
DAAG55--98--C--0041 and DAAD19--01--1--06.
AW is supported by the U.K.~Engineering and Physical Sciences Research Council.

\appendix

\section{Types and typical subspaces}
\label{app:types}
The following material can be found in most textbooks on information
theory, e.g.~\cite{cover:thomas,csiszar:koerner}, or in the original literature
on quantum information theory~\cite{schumacher:qdc,SW:coding,winter:qstrong}.
\par
For strings of length $n$ from a finite alphabet ${\cal X}$, which we generically
denote $x^n=x_1\ldots x_n\in{\cal X}^n$, we define the \emph{type of $x^n$} as
the empirical distribution of letters in $x^n$: i.e., $P$ is the type of $x^n$ if
$$\forall x\in{\cal C}\quad P(x)=\frac{1}{n}|\{ k:x_k=x \}|.$$
It is easy to see that the total number of types is upper bounded by
$(n+1)^{|{\cal X}|}$.
\par
The \emph{type class of $P$}, denoted ${\cal T}_P^n$, is defined as all
strings of length $n$ of type $P$. Obviously, the type class is obtained by
taking all permutations of an arbitrary string of type $P$.
\par
The following is an elementary property of the type class:
\begin{equation}
  \label{eq:type:cardinality}
  (n+1)^{-|{\cal X}|}\exp\bigl( nH(P) \bigr)
                        \leq |{\cal T}_P^n| \leq \exp\bigl( nH(P) \bigr),
\end{equation}
with the (Shannon) entropy $H(P)$.
\par
For $\delta>0$, and for an arbitary probability
distribution $P$, define the set of \emph{$P$--typical sequences} as
$${\cal T}^n_{P,\delta} := \left\{ x^n: \left|-\frac{1}{n}\log P^{\otimes n}(x^n)-H(P)\right|
                                                                                \leq \delta\right\}.$$
By the law of large numbers, for every $\epsilon>0$ and sufficiently large $n$,
\begin{equation}
  \label{eq:typical:prob}
  P^{\otimes n}({\cal T}^n_{P,\delta}) \geq 1-\epsilon.
\end{equation}
Furthermore:
\begin{align}
  \label{eq:typical:upper}
  |{\cal T}^n_{P,\delta}| &\leq \exp\bigl( n(H(P)+\delta) \bigr), \\
  \label{eq:typical:lower}
  |{\cal T}^n_{P,\delta}| &\geq (1-\epsilon)\exp\bigl( n(H(P)-\delta) \bigr).
\end{align}
\par
For a (classical) channel $W:{\cal X}\longrightarrow{\cal Y}$ (i.e. a stochastic map, taking
$x\in{\cal X}$ to a probability distribution $W_x$ on ${\cal Y}$) and a string
$x^n\in{\cal X}^n$ of type $P$ we denote the output distribution of $x^n$ in
$n$ independent uses of the channel by
$$W^n_{x^n} = W_{x_1}\otimes\cdots\otimes W_{x_n}.$$
Let $\delta>0$, and define the set of \emph{conditonal $W$--typical sequences} as
$${\cal T}^n_{W,\delta}(x^n) := \left\{ y^n: \left|-\frac{1}{n}\log W^n_{x^n}(y^n)-H(W|P)\right|
                                                                                \leq \delta\right\},$$
where $H(W|P)=\sum_x P(x)H(W_x)$ is the conditional entropy.
\par
Once more by the law of large numbers, for every $\epsilon$ and
sufficiently large $n$,
\begin{equation}
  \label{eq:c-typical:prob}
  W^n_{x^n}\bigl( {\cal T}^n_{W,\delta}(x^n) \bigr) \geq 1-\epsilon.
\end{equation}
Furthermore:
\begin{align}
  \label{eq:c-typical:upper}
  \bigl| {\cal T}^n_{W,\delta}(x^n) \bigr| &\leq \exp\bigl( n(H(W|P)+\delta) \bigr), \\
  \label{eq:c-typical:lower}
  \bigl |{\cal T}^n_{W,\delta}(x^n) \bigr| &\geq (1-\epsilon)\exp\bigl( n(H(W|P)-\delta) \bigr).
\end{align}
\par
All of these concepts and formulas have analogues as ``typical projectors'' $\Pi$ for
quantum state: by virtue of the spectral decomposition, the eigenvalues of a density
operator can be interpreted as a probability distribution over eigenstates. The subspaces
spanned by the typical eigenstates are the ``typical subspaces''. The trace of a density
operator with one of its typical projectors is then the probability of the corresponding
set of typical sequences.
\par
Notations like $\Pi^n_{\rho,\delta}$ etc. should be clear from this.
\par
There is only one such statement for density operators that we shall use, which is
not of this form:
\begin{lemma}[Operator law of large numbers]
  \label{lemma:law:large-numbers}
  Let $x^n\in{\cal X}^n$ be of type $P$, and let $W:{\cal X}\longrightarrow{\cal S}({\cal H})$
  be a cq--channel. Denote the average output state of $W$ under $P$ as
  $$\rho = \sum_x P(x) W_x.$$
  Then, for every $\epsilon>0$ and sufficiently large $n$,
  $$\tr\bigl( W^n_{x^n}\Pi^n_{\rho,\delta} \bigr)  \geq 1-\epsilon.$$
\end{lemma}
\begin{beweis}
  See~\cite{winter:qstrong}, Lemma 6.
\end{beweis}

\section{Miscellaneous facts}
\label{app:facts}
This appendix collects some standard facts about various functionals we use:
entropy, fidelity and trace norm.
\par
\begin{lemma}[Fannes~\cite{fannes}]
  \label{lemma:Fannes}
  Let $\rho$ and $\sigma$ be states on a $d$--dimensional Hilbert space, with
  $\| \rho-\sigma \|_1 \leq \delta$. Then
  $\bigl| H(\rho)-H(\sigma) \bigr| \leq \delta\log d + \tau(\delta)$, with
  \begin{equation*}
    \tau(\delta) = \begin{cases}
                     -\delta\log\delta & \text{ if }\delta\leq 1/4, \\
                     1/2               & \text{ otherwise.}
                   \end{cases}
  \end{equation*}
  Note that $\tau$ is a monotone and concave function.
  \qed
\end{lemma}
\par
\begin{lemma}[\cite{Fuchs:vdGraaf}]
  \label{lemma:norm-fidelity}
  Let $\rho$ and $\sigma$ be any two states on a Hilbert space. Then
  $$1-\sqrt{F(\rho,\sigma)} \leq \frac{1}{2}\| \rho-\sigma \|_1 \leq \sqrt{1-F(\rho,\sigma)}.$$
  \qed
\end{lemma}
\par
\begin{lemma}[Gentle measurement~\cite{winter:qstrong}]
  \label{lemma:gentle}
  Let $\rho$ be a (subnormalized) density operator, i.e. $\rho\geq 0$ and $\tr\rho\leq 1$,
  and let $0\leq X\leq \1$. Then, if $\tr(\rho X) \geq 1-\lambda$,
  $$\left\| \sqrt{X}\rho\sqrt{X}-\rho \right\|_1 \leq \sqrt{8\lambda}.$$
  \qed
\end{lemma}
\par
\begin{lemma}[Caratheodory's theorem~\cite{ziegler}, 1.6]
  \label{lemma:caratheodory}
  Let $v_1,\ldots,v_n$ be points in a $d$--dimensional $\R$--vector space,
  and let $p(1),\ldots,p(n)$ be probabilities (i.e., non--negative and
  summing to $1$. Then the convex combination
  $$v = \sum_{i=1}^n p(i) v_i$$
  can be expressed as a convex combination of (at most) $d+1$ of the $v_i$.
  \par
  As a consequence, there exist probability distributions $p_j$ on
  $\{1,\ldots,n\}$ and probability weights $\lambda_j$ such that
  for all $j$,
  $$v = \sum_{i=1}^n p_j(i) v_i, \quad \bigl| \supp\,p_j \bigr| \leq d+1.$$
  \qed
\end{lemma}

\section{Miscellaneous proofs}
\label{app:proofs}
\begin{beweis}[of proposition~\ref{prop:covering}]
  The proof follows closely the argument of~\cite{winterisation} and
  of~\cite{Ahlswede:Winter}: begin by constructing the typical projectors
  $\Pi_{x^n}$ of the $W_{x^n}$, which, for $x^n$ of type $P$,
  is defined as the sum of the eigenstate projectors of
  $W_{x^n}$ with eigenvalues in the interval
  $$\left[ \exp\bigl( -n(H(W|P)+\delta) \bigr) ; \exp\bigl( -n(H(W|P)-\delta) \bigr) \right],$$
  with the conditional entropy $H(W|P)=\sum_x P(x)H(W_x)$.
  For sufficiently large $n$, by the law of large numbers,
  $\tr(W_{x^n}\Pi_{x^n}) \geq 1-\epsilon$.
  Now define 
  $$\omega_{x^n} := \Pi\Pi_{x^n}W^n_{x^n}\Pi_{x^n}\Pi,$$
  where $\Pi$ is the typical projector of $\rho=\sum_x P(x)W_x$, i.e.
  the sum of the eigenstate projectors with eigenvalues in the interval
  $$\left[ \exp\bigl( -n(H(\rho)+\delta) \bigr) ; \exp\bigl( -n(H(\rho)-\delta) \bigr) \right].$$
  These concepts are taken from~\cite{schumacher:qdc} and~\cite{SW:coding},
  but see also appendix~\ref{app:types}.
  By the law of large numbers, for sufficiently large $n$,
  $$\tr(W_{x^n}\Pi) \geq 1-\epsilon^2/8.$$
  From this and the gentle measurement lemma~\ref{lemma:gentle}, we get
  \begin{equation}
    \label{eq:truncated}
    \bigl\| \omega_{x^n}-W_{x^n} \bigr\|_1 \leq 2\epsilon.
  \end{equation}
  The strategy is now to apply the operator Chernoff bound to the $\omega_{x^n}$:
  they are supported on a subspace of dimension $\leq \exp\bigl( n(H(\rho)+\delta) \bigr)$,
  and are all upper bounded by $\exp\bigl( -n(H(W|P)-\delta) \bigr)\Pi$.
  \par
  The only remaining obstacle is that we need a lower bound on
  $$\overline{\omega} = \frac{1}{|{\cal T}_P^n|}\sum_{x^n\in{\cal T}_P^n} \omega_{x^n}.$$
  To this end, let $\widehat\Pi$ be the projector onto the subspace spanned
  by eigenvectors of $\overline{\omega}$ with eigenvalues
  $\geq \exp\bigl( -n(H(\rho)-2\delta) \bigr)$. In this way, for sufficiently
  large $n$,
  $$\tr(\overline\omega\widehat\Pi) \geq 1-\epsilon.$$
  Defining the operators
  $$\widehat\omega_{x^n} := \widehat\Pi \omega_{x^n} \widehat\Pi,$$
  we can now apply lemma~\ref{lemma:op:chernoff} to the (rescaled)
  $\widehat\omega_{U^{(j)}}$, and get
  \begin{equation}\begin{split}
    \label{eq:there:it:is}
    \Pr&\left\{ \frac{1}{M}\sum_{j=1}^M \widehat\omega_{U^{(j)}}
                \not\in \bigl[(1\pm\epsilon)\widehat\Pi\overline\omega\widehat\Pi \bigr] \right\} \\
       &\phantom{:}
        \leq 2d^n\exp\left( -M\exp\bigl(-n(I(P;W)+3\delta)\bigr)\frac{\epsilon^2}{2\ln 2} \right)\!.
  \end{split}\end{equation}
  But
  $$\Omega := \frac{1}{M}\sum_{j=1}^M \widehat\omega_{U^{(j)}}
                \in \bigl[(1\pm\epsilon)\widehat\Pi\overline\omega\widehat\Pi \bigr]$$
  implies $\bigl\| \widehat\Pi(\Omega-\overline\omega)\widehat\Pi \bigr\|_1 \leq \epsilon$,
  which in turn implies
  \begin{equation}
    \label{eq:step-one}
    \bigl\| \widehat\Pi\Omega\widehat\Pi-\overline\omega \bigr\|_1 \leq 2\epsilon.
  \end{equation}
  In particular we get, invoking eq.~(\ref{eq:truncated}),
  $$\tr\Omega \geq \tr\overline\omega-2\epsilon
                                \geq 1-4\epsilon,$$
  Hence, with the gentle measurement lemma~\ref{lemma:gentle}
  appendix~\ref{app:facts}, we obtain
  \begin{equation}
    \label{eq:step-two}
    \bigl\| \widehat\Pi\Omega\widehat\Pi - \Omega \bigr\|_1 \leq \sqrt{32\epsilon}.
  \end{equation}
  Combining eqs.~(\ref{eq:step-one}) and (\ref{eq:step-two}) via the triangle
  inequality gives
  \begin{equation}
    \label{eq:step-three}
    \bigl\| \Omega - \overline\omega \bigr\|_1 \leq 2\epsilon+\sqrt{32\epsilon},
  \end{equation}
  and using eq.~(\ref{eq:truncated}) to replace $\omega_{x^n}$ by $W^n_{x^n}$ in both
  above operators, we get finally
  $$\left\| \frac{1}{M}\sum_j W^n_{U^{(j)}} - \sigma(P) \right\|_1 \leq 6\epsilon+\sqrt{32\epsilon}
                                                                   \leq 12\sqrt{\epsilon}.$$
  The complement of this event has probability smaller than
  eq.~(\ref{eq:there:it:is}), and since $\delta$
  was arbitrary we obtain our claim.
\end{beweis}
\par\bigskip
\begin{beweis}[of proposition~\ref{prop:HSW:type}]
  In~\cite{SW:coding}, it is proved that selecting
  $$N'=2(n+1)^{|{\cal X}|}\exp\bigl( n(I(P;W)-\delta) \bigr)$$
  codewords $u^{(i)}$ at random,
  i.i.d.~according to $P^{\otimes n}$, one can construct a
  canonical decoding POVM such that for the expectation (over the code ${\cal C}_{\rm HSW}$)
  of the average error probability, $p_E$, goes to zero:
  \begin{equation}
    \label{eq:SW:bound}
    \langle p_E \rangle_{{\cal C}_{\rm HSW}} \leq 9\epsilon+N'\exp\bigl( -n(I(P;W)-\delta/2) \bigr).
  \end{equation}
  (See~\cite{SW:coding}, eq.~(34).)
  The first thing we note is that (for sufficiently large $n$)
  $\epsilon=\exp(-\gamma n)$ for a constant $\gamma>0$
  depending on $\delta$: this follows by inspection of section III of~\cite{SW:coding},
  where $\epsilon$ is introduced as the loss of probability mass by removing
  non--typical contributions. As non--typicality is defined as large deviation
  events for a sum of independent random variables, of the form
  $$\log\lambda_{x^n} = \sum_k \log\lambda_{x_k},$$
  the Chernoff bound allows us to
  put exponential bounds on the non--typical mass.
  \par
  Hence eq.~(\ref{eq:SW:bound}) can be rewritten, for sufficiently large $n$,
  \begin{equation}
    \label{eq:exp-error}
    \langle p_E \rangle_{{\cal C}_{\rm HSW}} \leq \exp( -n\beta ),
  \end{equation}
  with some $\beta>0$.
  \par
  We want to show that ${\cal C}_{\rm HSW}\cap{\cal T}_P^n$ is a good approximation
  to a random code from the type class ${\cal T}_P^n$. Of course, it is not quite
  that, if only because it has a variable number of codewords!
  There is an easy fix to this problem: define, with
  $N=\exp\bigl( n(I(P;W)-\delta) \bigr)$,
  \begin{equation*}
    {\cal C}:=\text{First }N\text{ elements of }{\cal C}_{\rm HSW}\cap{\cal T}_P^n,
  \end{equation*}
  which makes sense because we can put the codewords in the order we select them.
  If the intersection is too small, define ${\cal C}$ to be empty.
  \par
  First of all, let us bound the error probability of ${\cal C}$:
  \begin{equation}\begin{split}
    \label{eq:participation}
    p_E({\cal C}) &\leq \frac{1}{N}\sum_{u^{(i)}\in{\cal T}_P^n}
                                                   \bigl( 1-\tr(W^n_{u^{(i)}})D_i \bigr) \\
                  &\leq \frac{1}{N}\sum_{i=1}^{N'} \bigl( 1-\tr(W^n_{u^{(i)}})D_i \bigr) \\
                  &=    \frac{N'}{N} p_E({\cal C}_{\rm HSW})
  \end{split}\end{equation}
  \par
  Now, that $|{\cal C}_{\rm HSW}\cap{\cal T}_P^n|<N$, happens extremely rarely: because
  $P^{\otimes n}({\cal T}_P^n) \geq (n+1)^{-|{\cal X}|}$, the expected cardinality
  of the intersection is larger than $2N$, for sufficiently large $n$. But then,
  using the Chernoff bound,
  \begin{equation*}\begin{split}
    \Pr\bigl\{ \bigl|{\cal C}_{\rm HSW}\cap{\cal T}_P^n\bigr| < N \bigr\}
            &\leq \exp\left( -N'\frac{1}{8\ln 2 (n+1)^{|{\cal X}|}} \right)  \\
            &\leq \exp(-N/4).
  \end{split}\end{equation*}
  By symmetry, it is clear that conditional on ${\cal C}\neq\emptyset$,
  the code ${\cal C}$ is a uniformly random code of $N$ words from ${\cal T}_P^n$,
  i.e. it can be described by i.i.d.~and uniformly picking codewords.
  \par
  Hence, denoting by $\widetilde{{\cal C}}$ a truly random code of $N$ words
  from ${\cal T}_P^n$, we have
  $$\frac{1}{2}\left\| {\rm Dist}({\cal C})-{\rm Dist}(\widetilde{{\cal C}}) \right\|_1
                                                                         \leq \exp(-N/4).$$
  Observe that the left hand side is the total variational distance of distributions.
  Thus, putting this together with eqs.(\ref{eq:participation}) and
  (\ref{eq:exp-error}), we obtain
  \begin{equation*}\begin{split}
    \langle p_E \rangle_{\widetilde{{\cal C}}}
        &\leq \langle p_E \rangle_{{\cal C}} + \exp(-N/4)   \\
        &\leq 2(n+1)^{|{\cal X}|}\exp(-n\beta) + \exp(-N/4) \\
        &\leq \exp(-n\beta/2),
  \end{split}\end{equation*}
  for sufficiently large $n$. But this in turn implies that
  $$\Pr\bigl\{ p_E(\widetilde{{\cal C}}) > \exp(-n\beta/4) \bigr\} \leq \exp( -n\beta/4 ),$$
  by the Markov inequality.
\end{beweis}
\par\bigskip
\begin{beweis}[of range bounds in theorem~\ref{thm:cqq:1way-key}]
  Here we prove that we may assume that $T$ is a deterministic function of $U$,
  $|T|\leq |{\cal X}|$ and $|U|\leq |{\cal X}|^2$:
  \par
  Observing
  \begin{equation*}\begin{split}
    I(U;B|T) - I(U;E|T) &= H(B|T)-H(B|UT) \\
                        &\phantom{=} - \bigl[ H(E|T)-H(E|UT) \bigr],
  \end{split}\end{equation*}
  we aim at writing the four conditional entropies on the right
  as averages over similar such quantities but with limited range of $U$ and $T$.
  To this end, observe that the channels $Q$ and $R$ induce a
  probability distribution $q(ut)$ on the values $ut$ of $\widetilde{U}:=UT$
  (of which $T$ clearly is a deterministic function),
  and that for each $ut$ there is the conditional distribution $P_{ut}$
  on ${\cal X}$:
  $$P_{ut}(x) = \Pr\{ X=x | U=u,T=t \},$$
  which has the property $\sum_{ut} q(ut) P_{ut} = P$.
  With these notations,
  \begin{align*}
    H(B|UT)     &= \sum_{ut} q(ut) S\bigl( \rho^B_{ut} \bigr),\text{ where} \\
    \rho^B_{ut} &= \sum_x P_{ut}(x)\rho^B_x,
  \end{align*}
  and similarly for $H(E|UT)$.
  \par
  For each $t$, let $q(t)=\sum_u q(ut)$, which allows us to write down
  the conditional distribution $q(\cdot|t)$ on the points $ut$:
  \begin{equation*}
    q(ut'|t) = \begin{cases}
                 \frac{1}{q(t)} q(ut) & \text{ for }t'=t, \\
                 0                    & \text{ otherwise.}
               \end{cases}
  \end{equation*}
  With this, the conditional distribution $P_t$ on ${\cal X}$ can be written
  \begin{equation*}
    P_t = \Pr\{ X=x | T=t \} = \sum_u q(u|t) P_{ut},
  \end{equation*}
  for which clearly $\sum_t q(t) P_{t} = P$.
  This allows us to write
  \begin{align*}
    H(B|T)   &= \sum_t q(t) S\bigl( \rho^B_t \bigr),\text{ where} \\
    \rho^B_t &= \sum_x P_t(x)\rho^B_x,
  \end{align*}
  and similarly for $E$.
  \par
  Now, invoking Caratheodory's theorem, lemma~\ref{lemma:caratheodory},
  we can write, for all $t$,
  \begin{equation}
    \label{eq:q-cond:decomp}
    q(\cdot|t) = \sum_j \lambda_{j|t}q_j(\cdot|t),
  \end{equation}
  with probabilities $\lambda_{j|t}$ and conditional distributions $q_j$
  such that for all $j$,
   \begin{equation}
    \label{eq:Pt:preserved}
    \sum_{ut'} q_j(ut'|t)P_{ut'} = P_t
  \end{equation}
  and $|\supp\,q_j(\cdot|t)| \leq |{\cal X}|$.
  Another application of Caratheodory's theorem gives a convex decomposition
  \begin{equation}
    \label{eq:q:decomp}
    q = \sum_k \mu_k q_k,
  \end{equation}
  such that the support of all the $q_k$ has cardinality $\leq |{\cal X}|$
  and for all $k$,
  \begin{equation}
    \label{eq:P:preserved}
    \sum_t q_k(t) P_t = P.  
  \end{equation}
  Eqs.~(\ref{eq:q-cond:decomp}) and~(\ref{eq:q:decomp}) define random
  variables $J$ and $K$, respectively: by eqs.~(\ref{eq:Pt:preserved})
  and~(\ref{eq:P:preserved}), for each value $JK=jk$ the conditional
  distribution of $T$ and $\widetilde{U}$ define variables
  $T_{jk}|\widetilde{U}_{jk}|X$, and so
  \begin{equation*}\begin{split}
    H(B|T) &- H(E|T) = H(B|TJ) - H(E|TJ) \\
           &\phantom{=}
                     = \sum_{jk} \Pr\{JK=jk\} \bigl[ H(B|T_{jk})-H(E|T_{jk}) \bigr].
  \end{split}\end{equation*}
  In the same way,
  \begin{equation*}\begin{split}
    -&H(B|UT) + H(E|UT) = - H(B|\widetilde{U}JK) + H(E|\widetilde{U}JK) \\
     &\phantom{====}    = \sum_{jk} \Pr\{JK=jk\}
                                    \bigl[ -H(B|\widetilde{U}_{jk})+H(E|\widetilde{U}_{jk}) \bigr].
  \end{split}\end{equation*}
  Hence there exist $j$ and $k$ such that
  $$I(U;B|T) - I(U;E|T) \leq I(\widetilde{U}_{jk};B|T_{jk}) - I(\widetilde{U}_{jk};E|T_{jk}),$$
  and $\widetilde{U}_{jk}$ and $T_{jk}$ satisfy the range constraints.
\end{beweis}
\par\bigskip
\begin{beweis}[of range bounds in theorem~\ref{thm:qqq:1way-key}]
  Here we prove that we may assume that $T$ is a deterministic function of $X$,
  $|T|\leq d_A^2$ and $|X|\leq d_A^4$:
  \par
  Denote the POVM elements of the measurement producing $x$ and $t$ by
  $P_{xt}$, and introduce the coarse--grained operators
  $P_t = \sum_x P_{xt}$. To decompose the POVM using convexity arguments,
  we rewrite the completeness conditions as
  \begin{align*}
    \frac{1}{d_A}\1 &= \sum_t \frac{\tr\,P_t}{d_A}\frac{P_t}{\tr\,P_t}
                         =: \sum_t q(t)\pi_t,                                       \\
    \pi_t           &= \sum_x \frac{\tr\,P_{xt}}{\tr\,P_t}\frac{P_{xt}}{\tr\,P_{xt}}
                         =: \sum_{xt'} q(xt'|t)\pi_{xt'}.
  \end{align*}
  Using Caratheodory's theorem, lemma~\ref{lemma:caratheodory}, we can write
  \begin{equation}
    \label{eq:OP:q:decomp}
    q = \sum_k \mu_k q_k,
  \end{equation}
  with distributions $q_k$ of support $\leq d_A^2$ and such that for all $k$,
  \begin{equation}
    \label{eq:OP:mm:preserved}
    \sum_t q_k(t) \pi_t = \frac{1}{d_A}\1.
  \end{equation}
  Using Caratheodory's theorem once more, we obtain, for each $t$, a decomposition
  \begin{equation}
    \label{eq:OP:q-cond:decomp}
    q(\cdot|t) = \sum_j \lambda_{j|t} q_j(\cdot|t),
  \end{equation}
  with conditional distributions $q_j(\cdot|t)$ of support $\leq d_A^2$
  and such that for all $j$,
  \begin{equation}
    \label{eq:OP:pi:preserved}
    \sum_{xt'} q_j(xt'|t) \pi_{xt'} = \pi_t.
  \end{equation}
  Now, let $\widetilde{X}:=XT$, which $T$ clearly is a function of.
  Then, eqs.~(\ref{eq:OP:q:decomp}) and~(\ref{eq:OP:q-cond:decomp}) define random
  variables $J$ and $K$, respectively: by eqs.~(\ref{eq:OP:mm:preserved})
  and~(\ref{eq:OP:pi:preserved}), for each value $JK=jk$ we have
  a POVM $P^{(jk)}$ (whose output variable we denote $\widetilde{X}_{jk}$
  the function $T$ of which we denote $T_{jk}$).
  Then (compare the previous proof),
  \begin{align*}
     &H(B|T)  - H(E|T)  = H(B|TJ) - H(E|TJ)                                             \\
     &\phantom{===}     = \sum_{jk} \Pr\{JK=jk\} \bigl[ H(B|T_{jk})-H(E|T_{jk}) \bigr]. \\
    -&H(B|XT) + H(E|XT) = - H(B|\widetilde{X}JK) + H(E|\widetilde{X}JK)                 \\
     &\phantom{===}     = \sum_{jk} \Pr\{JK=jk\}
                                    \bigl[ -H(B|\widetilde{X}_{jk})+H(E|\widetilde{X}_{jk}) \bigr].
  \end{align*}
  Hence there exist $j$ and $k$ such that
  $$I(X;B|T) - I(X;E|T) \leq I(\widetilde{X}_{jk};B|T_{jk}) - I(\widetilde{X}_{jk};E|T_{jk}),$$
  and $\widetilde{X}_{jk}$ and $T_{jk}$ satisfy the range constraints.
\end{beweis}

\end{document}